\pdfoutput=1
\documentclass[twocolumn,english,aps,superscriptaddress,prl,10pt]{revtex4-1}

\usepackage{babel}
\usepackage{amsmath}
\usepackage{amssymb}
\usepackage{graphicx}
\usepackage{physics}

\usepackage{comment}

\usepackage{epstopdf}

\usepackage{xcolor}
\usepackage{tikz}
\usetikzlibrary{decorations}
\usetikzlibrary{decorations.pathmorphing}
\usetikzlibrary{decorations.pathreplacing}
\usetikzlibrary{decorations.markings}
\usetikzlibrary{shapes.misc}
\usetikzlibrary{calc}

\newcommand{\appropto}{\mathrel{\vcenter{
  \offinterlineskip\halign{\hfil$##$\cr
    \propto\cr\noalign{\kern2pt}\sim\cr\noalign{\kern-2pt}}}}}

\begin{document}

\title{Thermal Ising Transition in the Spin-1/2 $J_1-J_2$ Heisenberg Model}

\author{Olivier Gauth\'e}
\email{\href{mailto:olivier.gauthe@epfl.ch}{olivier.gauthe@epfl.ch}}
\affiliation{Institute of Physics, Ecole Polytechnique F\'ed\'erale de Lausanne (EPFL), CH-1015 Lausanne, Switzerland}
\author{Fr\'ed\'eric Mila}
\affiliation{Institute of Physics, Ecole Polytechnique F\'ed\'erale de Lausanne (EPFL), CH-1015 Lausanne, Switzerland}

\date{\today}
\begin{abstract} 
Using an SU(2) invariant finite-temperature tensor network algorithm, we provide strong numerical evidence in favor of an Ising transition in the collinear phase of the spin-1/2 $J_1-J_2$ Heisenberg model on the square lattice. In units of $J_2$, the critical temperature reaches a maximal value of $T_c/J_2\simeq 0.18$ around $J_2/J_1\simeq 1.0$. It is strongly suppressed upon approaching the zero-temperature boundary of the collinear phase $J_2/J_1\simeq 0.6$, and it vanishes as $1/\log(J_2/J_1)$ in the large $J_2/J_1$ limit, as predicted by Chandra, Coleman and Larkin [Phys. Rev. Lett. 64, 88, 1990]. Enforcing the SU(2) symmetry is crucial to avoid the artifact of finite-temperature SU(2) symmetry breaking of U(1) algorithms, opening new perspectives in the investigation of the thermal properties of quantum Heisenberg antiferromagnets.
\end{abstract}

\maketitle

%%%%%%%%%%%%%%%%%%%%%%%%%%%%%%%%%%%% INTRODUCTION %%%%%%%%%%%%%%%%%%%%%%%%%%%%%%%%%%%%

The spin-1/2 Heisenberg model on the square lattice with nearest-neighbor coupling $J_1$ and next-nearest coupling $J_2$, also known as the $J_1-J_2$ model, has imposed itself as a paradigmatic model of frustrated quantum magnetism since Chandra and Dou\c{c}ot suggested in 1988 that it could host a quantum spin liquid phase around $J_2/J_1 = 1/2$~\cite{chandra_possible_1988}. This model, which is realized in various vanadium oxides~\cite{nath_magnetic_2008}, has also attracted a lot of attention recently as an effective model to describe the magnetic properties of iron-based superconductors~\cite{si_high-temperature_2016}.  After more than three decades of very intensive
theoretical investigation~\cite{gelfand_zero-temperature_1989, read_valence-bond_1989, dagotto_phase_1989, figueirido_exact_1990, bishop_phase_1998, jiang_spin_2012, hu_direct_2013, gong_plaquette_2014, poilblanc_quantum_2017, haghshenas_u1-symmetric_2018, yu_deconfinement_2018, wang_critical_2018, liu_gapless_2018, hasik_investigation_2021},
most of its properties are still debated. At zero temperature, the presence of an intermediate region with no magnetic long-range order between N\'eel order at small $J_2$ and collinear order at large $J_2$ is well accepted, but the physics in this intermediate range is still unsettled, with proposals ranging from a $\mathbb{Z}_2$ quantum spin liquid phase to a valence-bond crystal, and possibly even two intermediate phases~\cite{ferrari_gapless_2020, liu_gapless_2022, nomura_dirac-type_2021}. Regarding the thermal properties of the model~\cite{poilblanc_finite-temperature_2021, niggemann_frustrated_2021}, the main open question concerns the possibility of an Ising transition at finite temperature in the collinear phase, at which the system is expected to choose between the two helical states of pitch vectors $(0,\pi)$ and $(\pi,0)$. First predicted in 1990 by Chandra, Coleman, and Larkin (CCL)~\cite{chandra_ising_1990} on the basis of analytic arguments, direct numerical evidence for the spin-1/2 case has been impossible to obtain so far. The only case where direct numerical evidence could be obtained is that of classical spins, for which extensive Monte Carlo simulations have demonstrated the presence of a transition in the Ising universality class~\cite{weber_ising_2003}. For the spin-1/2 case, quantum Monte Carlo simulations cannot be used because they suffer from a very serious minus sign problem, and high-temperature series expansions have failed to detect a phase transition~\cite{singh_symmetry_2003, misguich_determination_2003}, leading to the suggestion that maybe the critical temperature is equal to zero because of quantum effects.  Building on the Monte Carlo results for classical spins and assuming that there is collinear order in the ground state, a self-consistent harmonic approximation has been used to include quantum fluctuations and come up with a prediction of the $J_2$ dependence of the critical temperature for various values of the spin~\cite{capriotti_ising_2004}. This is not a direct proof however since it relies on a semiclassical treatment of quantum fluctuations, and whether a transition is indeed present for spin-1/2 is still an unsolved issue. For iron-based superconductors, this is a very important one since this Ising transition might be at the origin of their electronic nematicity~\cite{si_high-temperature_2016}.

In this Letter, we address this problem with tensor network algorithms and come up with the first direct evidence of an Ising phase transition in the spin-1/2 $J_1-J_2$ model on the square lattice. The method relies on the representation of the density matrix as a purified quantum state using auxiliary degrees of freedom~\cite{verstraete_matrix_2004}, and on the explicit implementation of SU(2) symmetry during the imaginary time evolution to avoid the artifact of spontaneous SU(2) symmetry breaking, which is a systematic problem if the algorithm only respects the U(1) symmetry. With this algorithm, we have been able to identify a spontaneous breaking of the $C_{4v}$ symmetry using a corner transfer matrix renormalization group (CTMRG) algorithm, and to show that the transition is fully consistent with the 2D Ising universality class.

The spin-1/2 $J_1-J_2$ model on the square lattice is defined by the Hamiltonian
\begin{equation}
    \label{eq:model}
    \mathcal{H} = J_1 \sum_{NN} \textbf{S}_i \cdot \textbf{S}_j + J_2 \sum_{NNN} \textbf{S}_i \cdot \textbf{S}_j 
\end{equation}
where the components of $\textbf{S}_i$ are spin-1/2 operators, and where the sums over NN and NNN refer to pairs of nearest and next-nearest neighbors respectively.  We will concentrate on the case $J_1,J_2>0$.

%%%%%%%%%%%%%%%    METHOD %%%%%%%%%%%%%%%%%%%%%%%
Formally, the method relies on evaluating observables in a thermal ensemble defined by
\begin{eqnarray*}
    \label{eq:rho}
   \rho(\beta) = \Tr_\text{ancillas} \ket{\Psi(\beta)} \bra{\Psi(\beta)}
\end{eqnarray*} 
with  
\begin{eqnarray*}
   |\Psi(\beta)\rangle = e^{-\frac{1}{2}\beta \mathcal{H}} |\Psi(0)\rangle
\end{eqnarray*}
where each spin has an ancilla partner, and where $|\Psi(0)\rangle$ is a product of singlets between each spin and its ancilla partner~\cite{verstraete_matrix_2004}. At infinite temperature ($\beta=0$), all spin configurations are equally weighted after the trace over the ancilla degrees of freedom, while after the evolution in imaginary time to inverse temperature $\beta$, the trace over the ancilla leads to the canonical density operator $\rho(\beta)$. It is represented as a tensor product, and one proceeds in two steps: (i) calculation of the purified wave function $\ket{\Psi(\beta)}$ and (ii) contraction of the tensor network to calculate observables. 

For the purpose of identifying spontaneous symmetry breaking, it is of course crucial for $\Psi(\beta)$ to keep all symmetries of the problem. When $C_{4v}$ symmetry breaking occurs, $\Psi(\beta)$ should keep equal weight contributions for the two different symmetry sectors, and its elementary tensors should stay symmetric. It should only be through the CTMRG process that one sector is selected and $C_{4v}$ symmetry is spontaneously broken in the observables. However, in practice, step (i) can only be done approximately by keeping a finite bond dimension after each Trotter step of the imaginary time evolution, and this can lead to two types of artifacts: 

(a) $\Psi(\beta)$ explicitly breaks SU(2) symmetry -- weights inside a multiplet are different. This artifact must absolutely be avoided because it occurs at rather high temperature and is accompanied by a breaking of $C_{4v}$ symmetry, thus masking the transition we are looking for. To overcome it, we have modified the algorithm to implement SU(2) symmetry at the tensor level~\cite{singh_tensor_2012, schmoll_programming_2020}, preventing any symmetry breaking.

(b) $\Psi(\beta)$ is SU(2) symmetric but nevertheless explicitly breaks the $C_{4v}$ symmetry -- horizontal and vertical bonds are no longer equivalent. This is less of a problem because this artifact occurs at rather low temperature, and for all parameters for which we report results, the actual transition occurs above this artifact. 

We now briefly give some details about the exact algorithmic setup. More information can be found in the Supplemental Material~\footnote{See Supplemental Material for (i) a detailed explanation of the iPEPS algorithm; (ii) a benchmark with high temperature series expansion data~\cite{rosner_high-temperature_2003} and (iii) a thoughtful discussion of the simple update symmetry breaking artifact and its impact. The Supplemental Material includes 
Refs.~\cite{white_density_1992, wietek_thermodynamic_2019, jimenez_quantum_2021}}.
Our method is based on infinite projected entangled pair states (iPEPS)~\cite{verstraete_renormalization_2004} at finite temperatures~\cite{czarnik_projected_2012, czarnik_projected_2015, czarnik_time_2018, czarnik_time_2019}, whose accuracy is controlled by the bond dimension $D$. We used a next-nearest neighbor simple update~\cite{jiang_accurate_2008, corboz_simulation_2010} to apply imaginary time evolution on a $2\times 2$ unit cell.
While the value of the finite imaginary time steps may shift the temperature where explicit symmetry breaking occurs, it makes little change on the observables before this artifact.
With SU(2) symmetry implemented, $D$ cannot be set arbitrarily and must respect virtual space decomposition into SU(2) multiplets. The algorithm dynamically finds the most relevant symmetry sectors by keeping a fixed number of independent multiplets in the truncations. We observed that this decomposition does not depend on $J_2$ and pins $D$ to the values $D \in \{1, 4, 7, 8, 11, 16, 19, 22\}$.
To contract the tensor network and compute observables, we used the asymmetric CTMRG algorithm \cite{nishino_corner_1996, orus_simulation_2009, corboz_stripes_2011, corboz_competing_2014}.  Numerical precision is less crucial here than in the optimization part and we only implemented the less technical U(1) symmetry~\cite{bauer_implementing_2011, singh_tensor_2011}. The accuracy of the contraction is controlled by the corner dimension $\chi$.

\begin{figure}[ht!]
\includegraphics[width=0.48\textwidth]{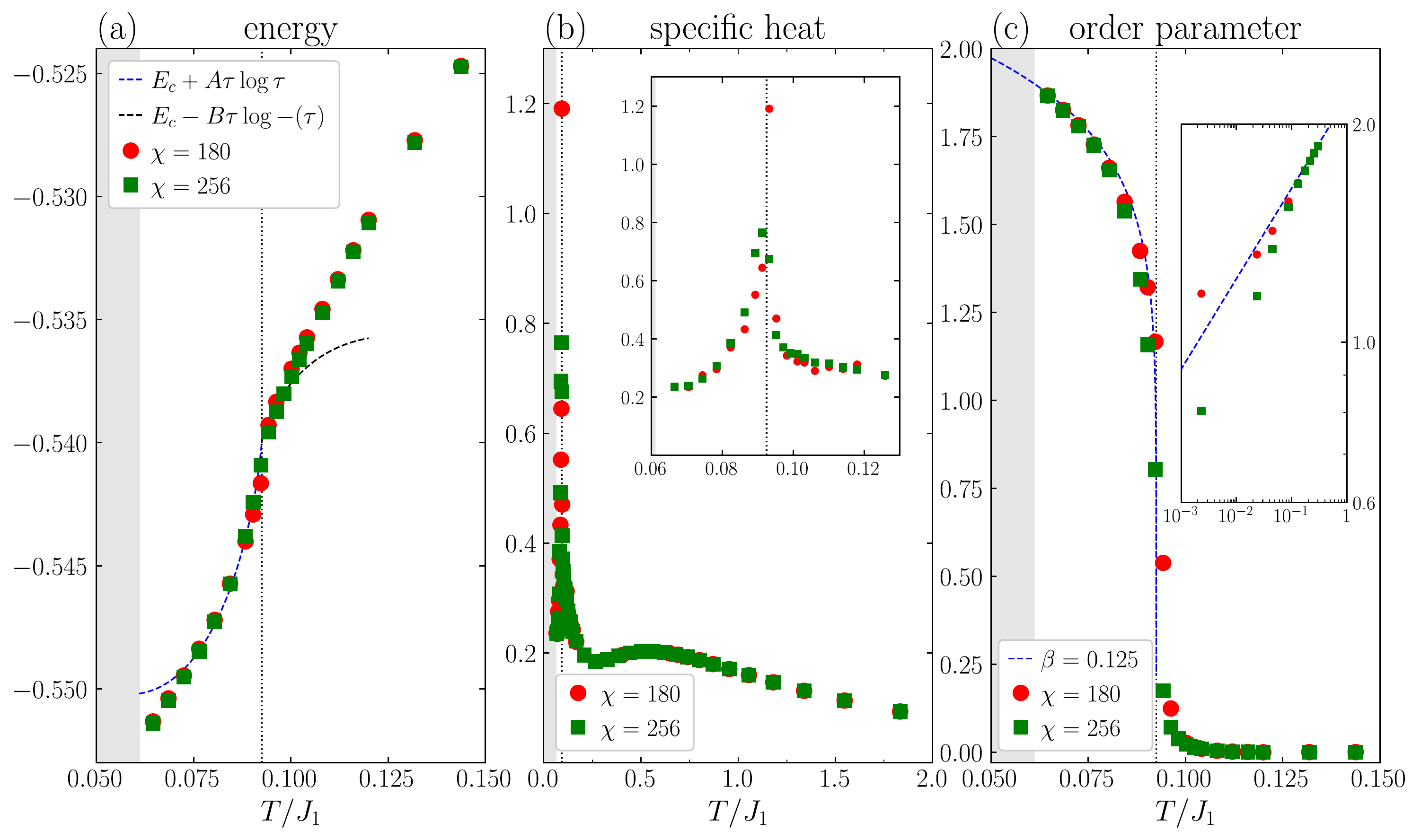}
\caption{\footnotesize{Various observables versus temperature for $J_2 = 0.85$  with $D=16$. The black dotted line marks the estimated critical temperature $T_c/J_1 = 0.093$. The grey area denotes the range below which the simple update has an unphysical artifact.
(a) Energy per site. The fit corresponds to a critical exponent $\alpha=0$ with reduced temperature $\tau = 1 - T/T_c$;
(b) Specific heat. Inset: enlargement around $T_c$;
(c) Order parameter $|\sigma|$ (see Eq.~\ref{eq:sigma}). It can be very well fitted with an Ising critical exponent $\beta=1/8$. The inset displays the same data in a semi-log scale in $\tau$.}}
\label{fig:observables}
\end{figure}

%%%%%%%%%%%%%%%%%%%%%%  The observables   %%%%%%%%%%%%%%%%%%%%%%%

We compute the reduced density matrices for nearest and next-nearest neighbor pairs, which allows us to extract the mean energy per site. We have benchmarked our results with high-temperature series expansion results \cite{rosner_high-temperature_2003} and found perfect agreement at high temperature (see the Supplemental Material). The specific heat is then obtained by numerical derivative of the energy. For the order parameter associated with $C_{4v}$ symmetry breaking, we have chosen the (non-normalized) difference between vertical and horizontal nearest-neighbor bonds inside the unit cell:
\begin{equation}
%\sigma = \sum_{\langle i,j \rangle_\vert} \textbf{S}_i \cdot \textbf{S}_j -  \sum_{\langle i,j \rangle_-} \textbf{S}_i \cdot \textbf{S}_j.
\sigma = \sum_{\langle i,j \rangle,v} \textbf{S}_i \cdot \textbf{S}_j -  \sum_{\langle i,j \rangle,h} \textbf{S}_i \cdot \textbf{S}_j.
\label{eq:sigma}
\end{equation}

Several correlation lengths $\xi_i$ can be extracted from the CTMRG-approximated transfer matrix eigenvalues according to: $
1/\xi_i = \ln \abs{\lambda_1 / \lambda_i}$, where $\lambda_i$ is the $i$th largest eigenvalue.
The degeneracy of the eigenvalues can be used to classify these correlation lengths according to SU(2) representations. This allows one in particular to distinguish the magnetic correlation length, a triplet diverging at zero temperature, from the Ising correlation length, a singlet diverging at the critical temperature.

\begin{figure}[ht!]
\includegraphics[width=0.48\textwidth]{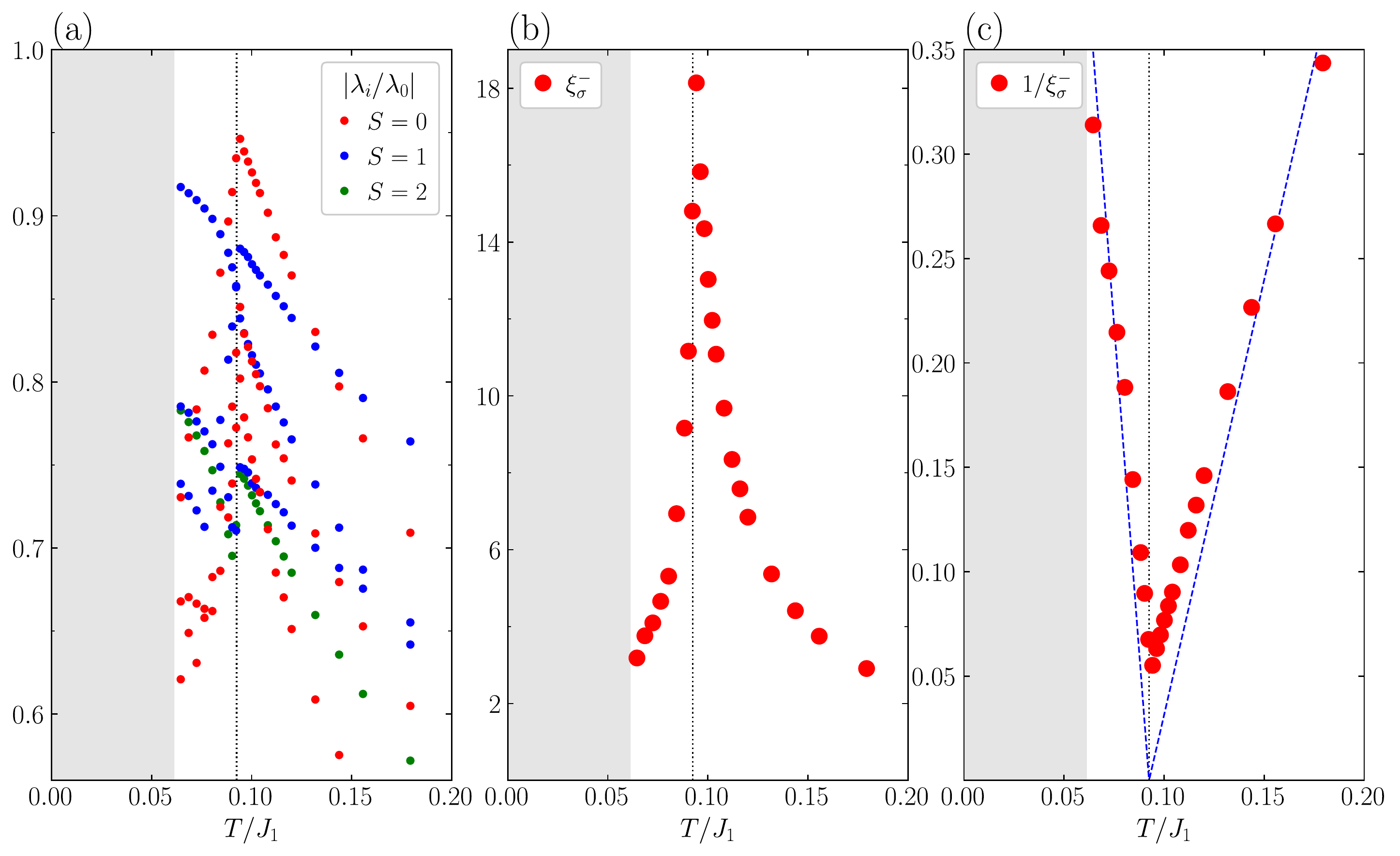}
\caption{\footnotesize{Spectrum of transfer matrix and correlation length for $J_2 = 0.85$ with $D=16$ and $\chi = 256$. Beyond the phase transition, horizontal and vertical lengths differ, and we plot only the smallest one for readability. The grey area denotes the range below which the simple update has an unphysical artifact. (a) Eigenvalues of the transfer matrix  $\lambda_i/\lambda_0$. The multiple level crossings are typical of a second order phase transition. The singlet that becomes the largest eigenvalue in a narrow temperature range corresponds to the correlation length $\xi_\sigma$ of the Ising order parameter. (b) Largest singlet correlation length $\xi_\sigma$. Close to the transition, its value is bounded by finite-$\chi$ effects. (c) Inverse of $\xi_\sigma$, showing the compatibility with $\nu=1$.}}
\label{fig:transf_spec}
\end{figure}

%%%%%%%%%%%%%%%%%%   J2 = 0.85   %%%%%%%%%%%%%%%%%%%%%%%%%%%%%%%%%%%
Let us start the presentation of the results by a thorough discussion of the case $J_2/J_1=0.85$.
For $D \leq 11$, the artifact of the $C_{4v}$ symmetry breaking of $\Psi(\beta)$ during imaginary time evolution occurs at a fairly high temperature, and we did not find any evidence of a phase transition above it. However, for $D=16$, we observe clear signs of a phase transition at a temperature $T_c/J_1 \simeq 0.093$, 
as shown in Figs.~\ref{fig:observables} and \ref{fig:transf_spec}: (i) the energy has a singularity; (ii) the specific heat has a very narrow peak at  $T/J_1 \simeq 0.093$, below a broad maximum at a higher temperature typical of antiferromagnets; (iii) the order parameter takes off very abruptly; and (iv) the correlation length diverges on both sides of the transition.

This phase transition is fully compatible with the 2D Ising universality class~\cite{baxter_square-lattice_1985}. First of all, the diverging correlation length corresponds to a nondegenerate eigenvalue, in agreement with the scalar order parameter of Eq. (\ref{eq:sigma}). The development of this order parameter is very steep, consistent with a small exponent $\beta$, and assuming $\beta=1/8$ leads to a critical temperature $T_c/J_1 \simeq 0.093$ that is compatible with the peak of the specific heat and the divergence of the correlation length. With this critical temperature, the exponent of the correlation length measured not too close to the critical temperature is consistent with $\nu=1$ (very close to the phase transition, the critical behavior is sensitive to the finite value of $\chi$). 
Finally, the behavior of the energy close to the transition is compatible with $T\log(T_c-T)$, in agreement with the expected logarithmic divergence of the specific heat ($\alpha=0$).

Interestingly, the largest triplet eigenvalue of the transfer matrix, which is only smaller than the next-to-leading singlet eigenvalue in a narrow parameter range, and the associated correlation length, which governs the decay of the spin-spin correlation function, continues to grow at low temperature, a behavior consistent with the expected divergence at zero temperature for a 2D antiferromagnet with long-range order.

\begin{figure}[ht!]
\includegraphics[width=0.48\textwidth]{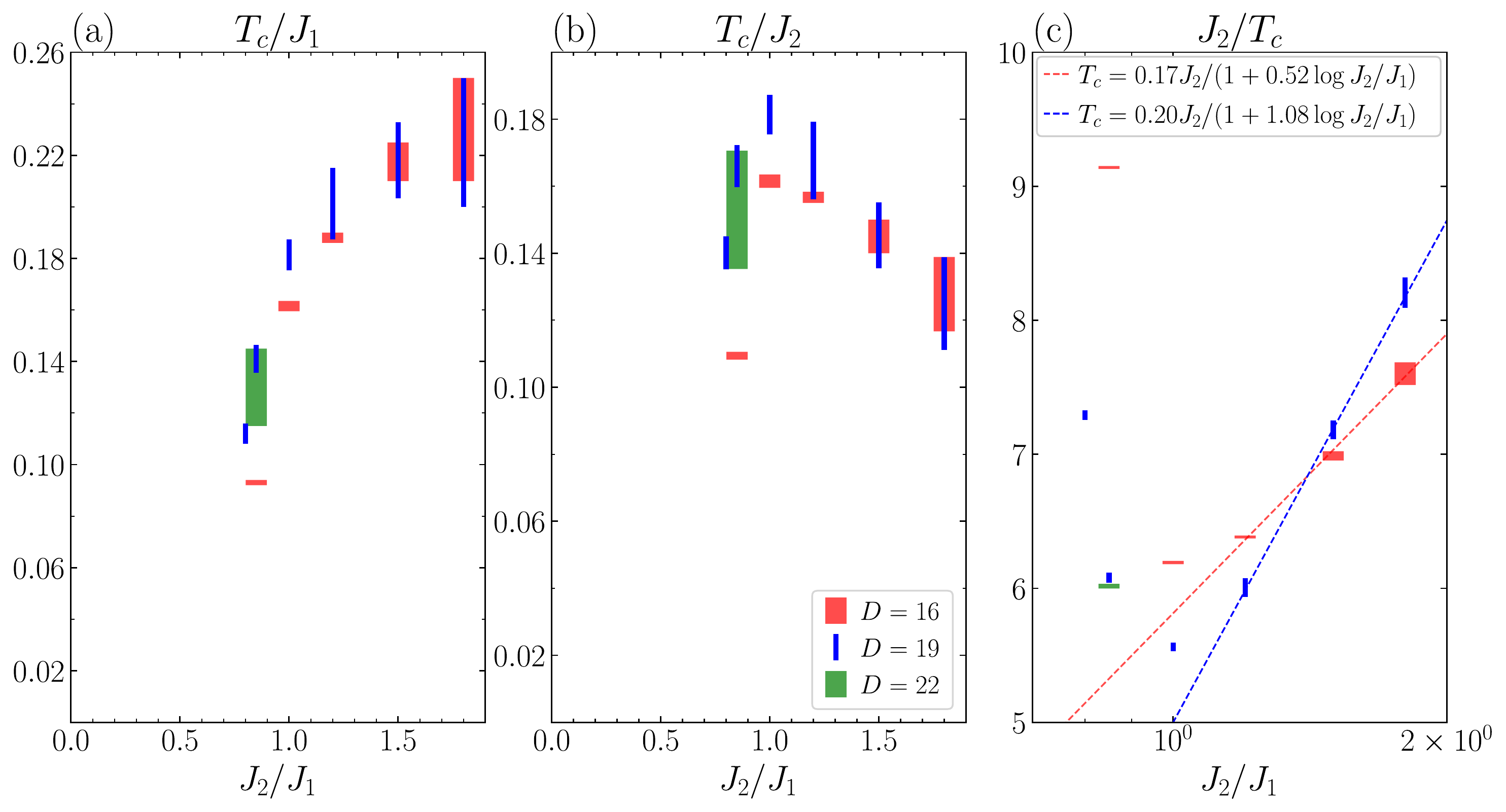}
\caption{\footnotesize{Phase diagram of the $J_1-J_2$ model. Left: Critical temperature as a function of $J_2/J_1$. Our results are consistent with a critical temperature vanishing at the quantum critical point $J_{2c}/J_1 \sim 0.6$. Center: Critical temperature in units of $J_2$. The curve reaches a maximum before going to 0 for $J_2 \rightarrow \infty$. Right: test for the analytical prediction $T_c = a J_2 / (1 + b \log J_2/J_1) $ in the large $J_2/J_1$ limit.}}
\label{fig:phase_diagram}
\end{figure}

%%%%%%%%%%%%%%%%%%%   Phase diagram    %%%%%%%%%%%%%%%%%%%%%%
The same analysis can be extended to larger values of $J_2/J_1$ up to $J_2/J_1=1.80$. For $J_2/J_1 \geq 2$, no transition could be detected before the artifact of the $C_{4v}$ symmetry breaking of $\Psi(\beta)$ occurs. The results for $T_c/J_1$ and $T_c/J_2$ are plotted in Figs.~\ref{fig:phase_diagram}(a) and \ref{fig:phase_diagram}(b) respectively. At fixed $\chi$, the finite corner dimension leads to an overestimation of the critical temperature: indeed a finite $\chi$ imposes a finite effective correlation length $\xi(\chi) < \xi(\infty)$. Accordingly, in the region where the infinite system is still disordered, but $\xi(\chi)$ is significantly smaller than the physical correlation length $\xi(\infty)$, hence smaller than the typical size of ordered domains, we measure a nonzero order parameter which then vanishes when using a larger $\chi$. Away from the transition, only small values of $\chi$ are needed for the observables to converge [see Fig. \ref{fig:observables}(c)]. Hence we set our upper bound as the lowest temperature where $\sigma=0$ for our largest $\chi$ and the lower bound as the highest temperature where $|\sigma|>0$ is converged in $\chi$.

While $D=16$ is the smallest bond dimension for which we observe a phase transition, we also considered larger bond dimensions. For large $J_2/J_1$, the dependence on $D$ is smaller than the error bars due to the finite value of $\chi$. For smaller values of $J_2$, the results change significantly between $D=16$ and $D=19$, and $T_c$ systematically increases from $D=16$ to $D=19$, a strong indication that it does not vanish in the infinite bond dimension limit. In addition, we have been able to converge a small number of points for the very challenging bond dimension $D=22$. These results sit between $D=16$ and $D=19$, and very close to the $D=19$ case. Therefore we believe that our results at $D=19$ give a reasonably accurate quantitative estimation of the critical temperature of the model. 

The phase diagram calls for a few general remarks. First, the critical temperature appears to go to zero at the critical ratio where collinear order sets in, as in the classical case. In the opposite limit of large $J_2/J_1$, $T_c/J_2$ also decreases, and, as we shall see, the behavior is consistent with CCL, whose theory predicts that $T_c/J_2$ vanishes as $1/\log (J_2/J_1)$. The maximum occurs around $J_2/J_1=1.0$, with $T_c/J_2\simeq 0.18$.

At intermediate values of $J_2/J_1$, the overall behavior of $T_c/J_2$ as a function of $J_2/J_1$ agrees qualitatively with the analysis of Capriotti et al., with a flat maximum around $J_2/J_1 \simeq 1$, and a slow decrease at larger $J_2/J_1$, as predicted by CCL. The maximal value of $T_c/J_2\simeq 0.18$ is smaller than that of Capriotti et al., $T_c/J_2\simeq 0.24$, but considering the nature of the approach of Capriotti et al., which is semiclassical in essence, such a semiquantitative agreement for spin-1/2 is very satisfactory.

In the large $J_2/J_1$ limit, CCL's approach predicts that the critical temperature decreases asymptotically as $T_c/J_2 = a J_2/ (1 + b \log J_2/J_1)$.  We tested this prediction in Fig.~\ref{fig:phase_diagram}(c), and the three largest values of $J_2$ are already consistent with this semilog asymptotic behavior, with $a=0.17,b=0.52$ for $D=16$ and $a=0.20,b=1.08$ for $D=19$. Note that this regime was not accessible in the investigation of the classical case by Weber et al.~\cite{weber_ising_2003} because $b=0.135$ is much smaller in that case, and logarithmic corrections would only be visible for values of $J_2$ much larger than $J_2/J_1=2$, the largest value for which an Ising transition could be detected. So the present results constitute to the best of our knowledge the first numerical confirmation of the asymptotic behavior at large $J_2$ predicted in CCL. Note that our values of $b$ are consistent with the prediction based on CCL by Weber et al, $b=0.78$, while our value for the overall slope $a$ is significantly smaller than the estimate based on CCL, $a=0.496$, a trend already observed for the classical case and attributed to the lack of quantitative information on the actual height of the energy barrier to go from one domain to the other.

Let us now discuss in more detail the small $J_2$ case.
All zero temperature simulations~\cite{wang_critical_2018, ferrari_gapless_2020, nomura_dirac-type_2021, liu_gapless_2022} point to a transition from a gapped phase (a $\mathbb{Z}_2$ spin liquid or a valence-bond solid phase) to the collinear phase around $J_2/J_1=0.6$. This is roughly consistent with our numerical results if we assume that $T_c$ vanishes linearly upon reducing $J_2/J_1$: a linear extrapolation of the last two points crosses the horizontal axis at $J_2/J_1\simeq 0.6$. However, our results would be hard to reconcile with the zero-temperature results if, as in the classical case in Ref.~\cite{weber_ising_2003}, $T_c$ was vanishing as a square root, with a vertical slope. Such a behavior would only be consistent with a critical temperature vanishing at a much larger value of $J_2/J_1$, of the order of 0.75. However, the argument put forward by Weber et al to explain the square root behavior does not apply to the quantum case. In the classical case, the collinear phase is in competition with the N\'eel phase at finite temperature, and the N\'eel phase is favored by thermal fluctuations, leading to a cross-over temperature that bends toward the collinear phase. This cross-over temperature grows as $\sqrt{J_2/J_1-1/2}$ and sets an upper bound to the Ising temperature, which was found  numerically to follow the same behavior. In the spin-1/2 case, by contrast, the collinear phase is believed to be in competition with a gapped phase (be it a $\mathbb{Z}_2$ phase or a valence-bond solid phase), and thermal fluctuations are expected to favor the collinear phase since the entropy grows as a power law in an ordered phase but is exponentially small in a gapped phase. So the cross-over temperature is expected to bend toward the gapped phase and cannot serve as an upper bound to the Ising transition. 

An alternative explanation is that the vanishing of the critical temperature at the boundary of the collinear phase is related to the vanishing of the spin stiffness in the collinear phase. Indeed,  exact diagonalizations~\cite{einarsson_direct_1995} and Schwinger bosons~\cite{trumper_schwinger-boson_1997} both point to a rapid but continuous and linear suppression of the spin stiffness around $J_2/J_1=0.6--0.65$ for the spin-1/2 case. Now, the energy scale of the effective Ising model that would describe this transition is set by the energy of a domain wall between two collinear domains with wave vectors $(0,\pi)$ and $(\pi,0)$ respectively, and this energy is expected to vanish if the stiffness vanishes. So, our results can be explained by a vanishing stiffness. Note however that we have not been able to get results at smaller values of $J_2/J_1$ because the CTMRG algorithm stops converging at low temperature for $J_2/J_1=0.75$. Further improvements (if at all possible) would be necessary to get reliable results in that range. In any case, the scenario put forward by Capriotti et al~\cite{capriotti_ising_2004}, with a critical temperature vanishing at $J_2/J_1 \simeq 0.6$ as a square root, as in the classical case, is not supported by our results. 

%%%%%%%%%%%%%%%%%%%%%%%%   Conclusion %%%%%%%%%%%%%%%%%%%%%%%%%%
To summarize, using the finite-temperature version of iPEPS, we have provided the first unambiguous and direct evidence of a thermal Ising transition in the collinear phase of the spin-1/2 $J_1-J_2$ model on the square lattice. It corresponds to the spontaneous breaking of the $C_{4v}$ symmetry, and the Ising 2D universality class has been demonstrated by a careful analysis of the order parameter, the correlation length, the energy, and the specific heat. 
Although limited, the range of values we could study, $0.8\le J_2/J_1 \le 1.8$, turned out to be enough to probe the small $J_2/J_1$ regime, with evidence that $T_c$ goes to zero linearly when $J_2/J_1$ approaches 0.6, the intermediate regime, with a maximum of $T_c/J_2=0.18$ around $J_2/J_1=1.0$, and the large $J_2/J_1$ regime, where we have been able to confirm the prediction of Chandra, Coleman and Larkin that $T_c/J_2$ vanishes as $1/\log(J_2/J_1)$. 

Beyond the $J_1/J_2$ model, we note that our approach relies on the implementation of the full SU(2) symmetry, which turned out to be crucial to obtain valid results. With the SU(2) symmetry implemented, this algorithm proved to be very robust and to give access to a large part of the phase diagram down to very low temperatures, previously out of reach by other methods. This opens the way to a systematic investigation of the thermal properties of frustrated quantum Heisenberg antiferromagnets, and more generally of strongly correlated systems for which quantum Monte Carlo simulations suffer from a severe minus sign problem.
\\

\begin{acknowledgments}
We acknowledge very useful discussions with Philippe Corboz and Andreas L\"auchli.
This work has been supported by the Swiss National Science Foundation. O.G. thanks Sylvain Capponi and Didier Poilblanc for insightful discussions and Fabien Alet, Juraj Hasik and Lo\"ic Herviou for advice with the code.
\end{acknowledgments}

\bibliographystyle{apsrev4-1}
\bibliography{heisenberg_J1J2}

\end{document}

%%%%%%%%%%%%%%%%%%%%%%%%%%%%%%%%%%%%%%%%%%%%%%%%%%%%%%

\title{Supplemental Material for:\\
Thermal Ising transition in the spin-1/2 $J_1-J_2$ Heisenberg model}
%%%%%%%%%%%%%%%%%%%%%%%%%%%%%%%%%%%%%%%%%%%%%%%%%%%%%%

\author{Olivier Gauth\'e}
\affiliation{Institute of Physics, Ecole Polytechnique F\'ed\'erale de Lausanne (EPFL), CH-1015 Lausanne, Switzerland}
\author{Fr\'ed\'eric Mila}
\affiliation{Institute of Physics, Ecole Polytechnique F\'ed\'erale de Lausanne (EPFL), CH-1015 Lausanne, Switzerland}
%%%%%%%%%%%%%%%%%%%%%%%%%%%%%%%%%%%%%%%%%%%%%%%%%%%%%%

\date{\today}
\maketitle

%%%%%%%%%%%%%%%%%%%%%%%%%%%%%%%%%%%%%%%%%%%%%%%%%%%%%%
\section{implementation details}
In this section, we provided details about the exact setup we used to produce our results. The source code can be made available upon request.

%%%%%%%%%%%%%%%%%%%%%%%%%%%%%%%%%%%%%%%%%%%%%%%%%%%%%%
\subsection{iPEPS ansatz}
Projected entangled pair states (PEPS) are the generalization of the 1-dimensional Density Matrix Renormalization group~\cite{white_density_1992} to higher dimensions~\cite{verstraete_renormalization_2004}.
They do not suffer from the Quantum Monte Carlo sign problem and can therefore be applied to frustrated magnets such as the $J_1-J_2$ model.

Beyond wavefunctions, they can be used to describe thermal ensembles, the most common one being known as the purification process~\cite{verstraete_matrix_2004}: the thermal ensemble is represented by a purified wavefunction living in an extended Hilbert space by adding an auxiliary Hilbert space acting as a thermal bath. To recover observables, one traces out the auxiliary space degrees of freedom. In practice, this auxiliary space is taken into account by adding an ancilla leg to local tensors with the same dimension as the physical variable.

\begin{figure}
	\centering
	\includegraphics[width=0.4\linewidth]{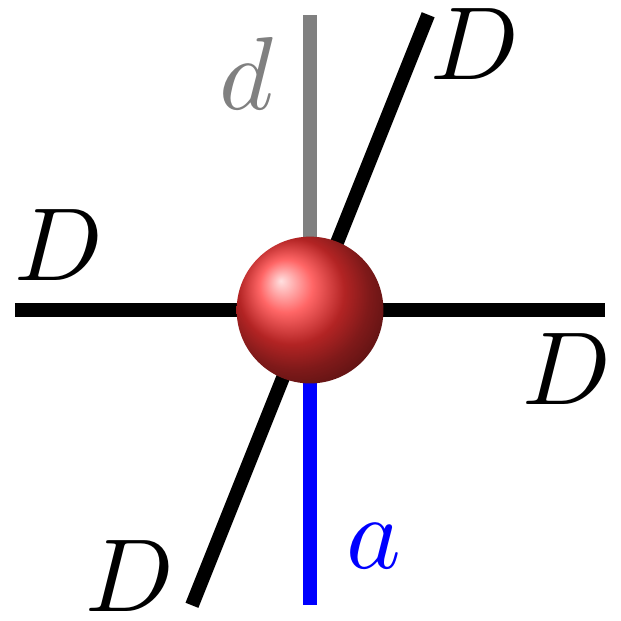}
	\caption{\footnotesize{Local rank-6 tensor, with physical leg with dimension $d$ (grey), ancilla leg with dimension $a$ (blue) and four virtual legs with dimension $D$ (black). Throughout we take $a=d=2$.}}
	\label{fig:tensor}
	\includegraphics[width=1.0\linewidth]{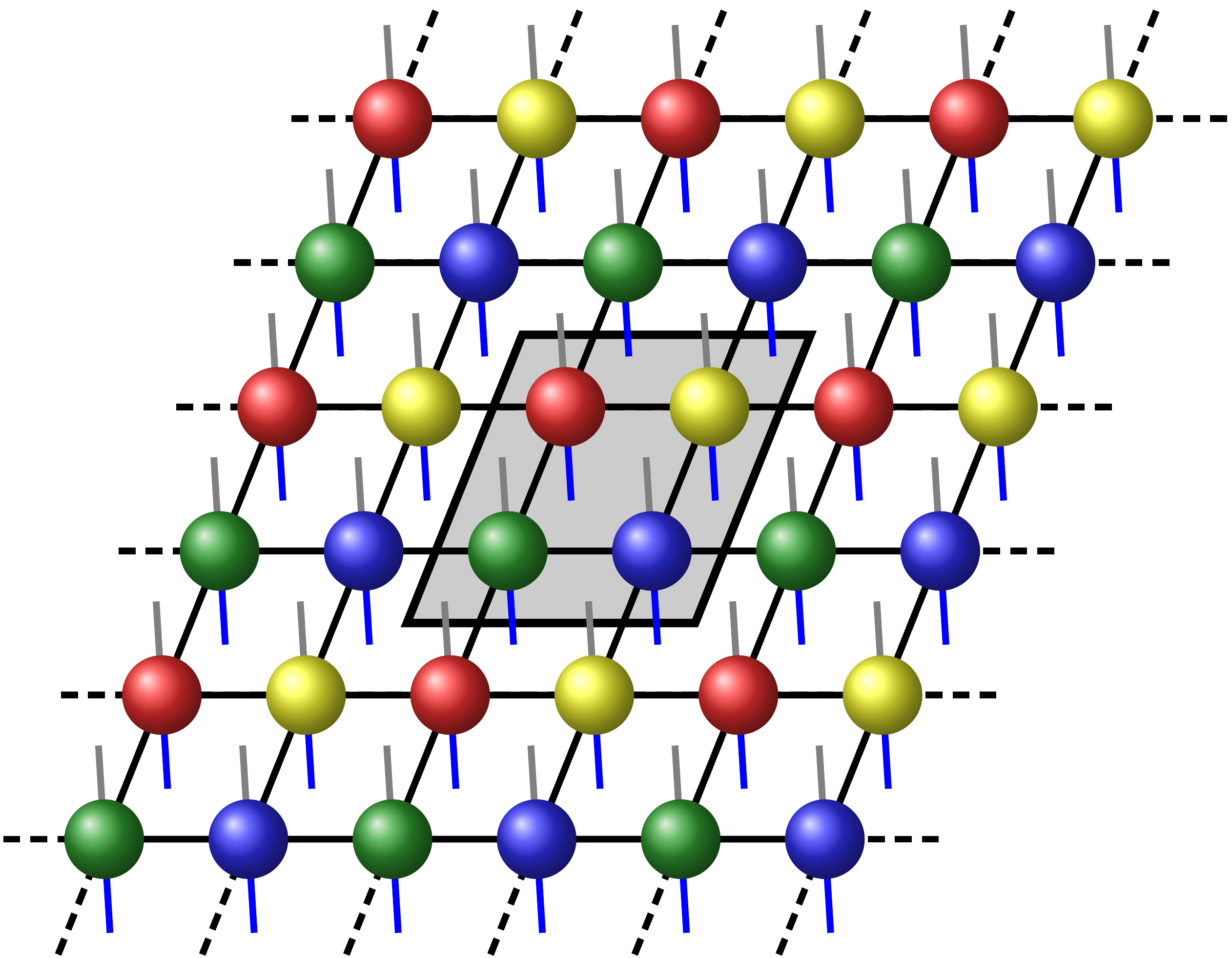}
	\caption{\footnotesize{The 2x2 unit cell used to tile the square lattice.}}
	\label{fig:iPEPS}
\end{figure}

Infinite PEPS (iPEPS) allows one to obtain results directly in the thermodynamic limit by repeating a fixed unit cell in all directions to tile the lattice.  Inside the unit cell, each site is described by a local rank-6 tensor of shape $(d,a,D,D,D,D)$, where the axes respectively correspond to the physical spin, the ancilla and the up, right, down and left bonds with a virtual dimension $D$ taken as large as possible (see Fig.~\ref{fig:tensor}). For a spin-$1/2$ thermal ensemble, $d=a=2$. We used a $2\times2$ site unit cell with 8 non-equivalent bonds, as required by next nearest neighbor simple update (see Fig.~\ref{fig:iPEPS}). In the contraction process, a bilayer tensor network with only virtual bonds is defined by contracting physical and ancilla legs of the bra and ket layers.

%%%%%%%%%%%%%%%%%%%%%%%%%%%%%%%%%%%%%%%%%%%%%%%%%%%%%%
\subsection{Simple update}
To simulate a thermal ensemble, the system is initialized at the exact infinite temperature product state, then one applies imaginary time evolution to reach a given temperature. 

Here, we use the simple update algorithm~\cite{jiang_accurate_2008, corboz_simulation_2010} for imaginary time evolution, an algorithm that has been successfully applied to the Shastry Sutherland model~\cite{wietek_thermodynamic_2019, jimenez_quantum_2021}. In this scheme, the Hamiltonian is decomposed into its elementary terms acting on the unit cell. Then, using a small imaginary time step $\tau$ the time evolution operator is decomposed into Trotter-Suzuki gates. We used the second-order Trotter-Suzuki decomposition by reversing the gate order at every other step.

\begin{figure}
	\centering
	\includegraphics[width=1.0\linewidth]{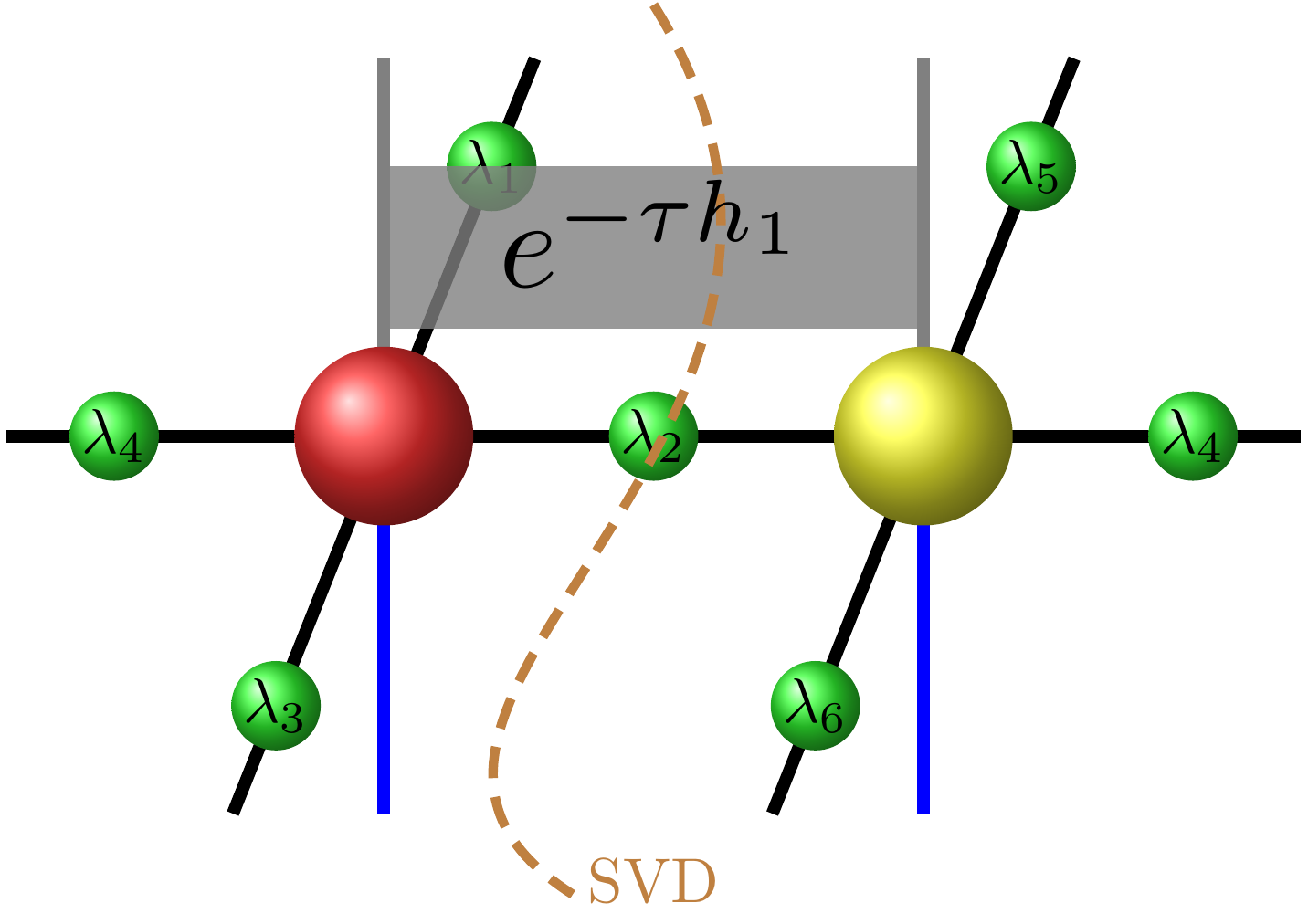}
	\caption{\footnotesize{First neighbor simple update. A gate $\exp(-\tau h_1)$ is applied to two first neighbor tensors. In the simple update algorithm, diagonal weights $\lambda_b$ are added to diagonal bonds to simulate an infinite environment. The updated bond is then renormalized using an SVD, generating a new set of weights $\lambda$ associated with this bond.}}
	\label{fig:SU_1st}
	\includegraphics[width=1.0\linewidth]{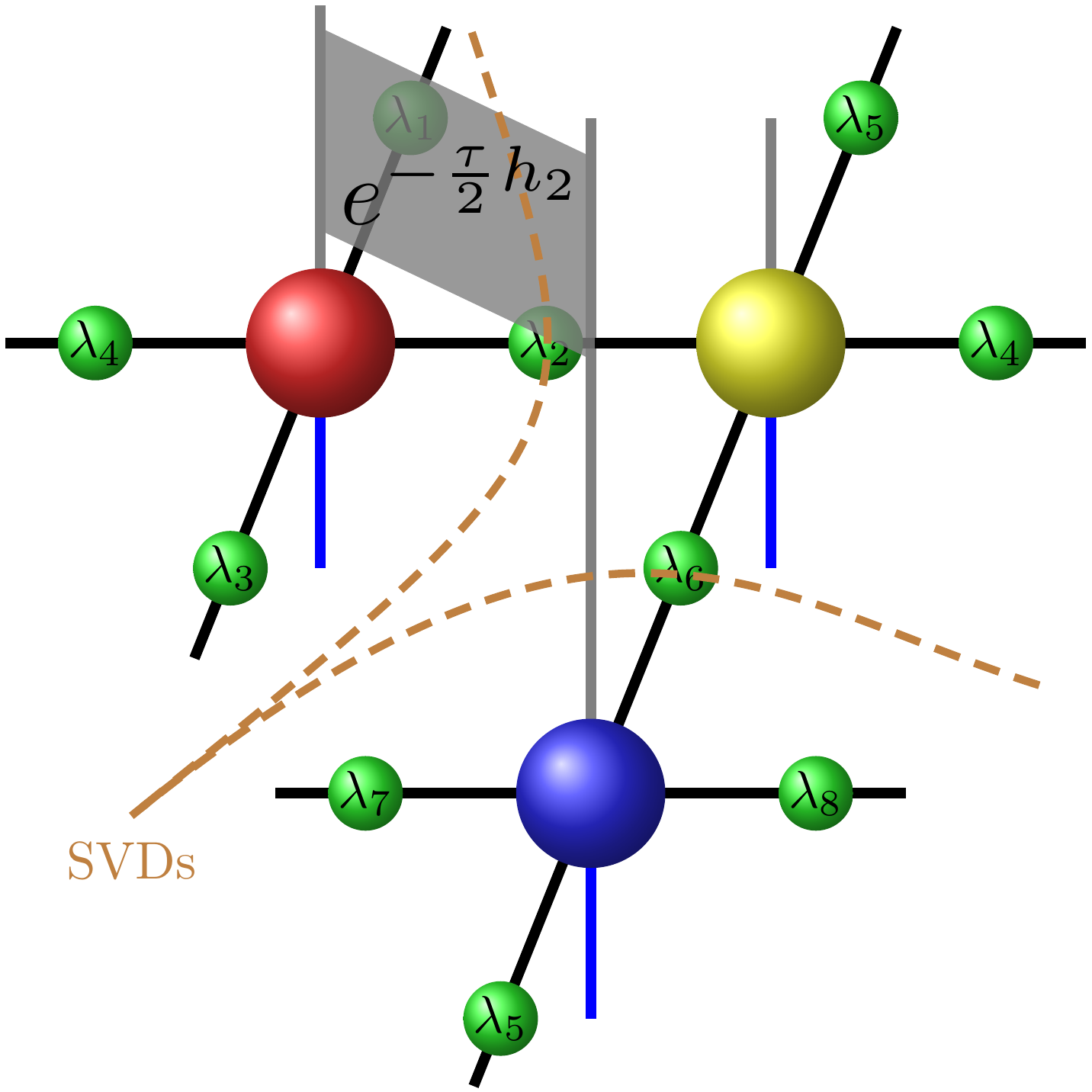}
	\caption{\footnotesize{Second neighbor simple update. One of the nearest-neighbor sites is used as an intermediate tensor (yellow) to link the two next-nearest neighbor sites and apply the associated gate. The two bonds connecting this intermediate site with the two next-nearest neighbor sites are then renormalized successively with two SVDs, yielding 3 updated tensors. One time step is executed by applying twice $\exp(-\tau h_2/2)$ with the two possible choices of nearest neighbor site for the intermediate tensor.}}
	\label{fig:SU_2nd}
\end{figure}

For each elementary term, site tensors are contracted with the associated gate, then bond renormalization is required to keep the bond dimension constant. The renormalization aims to find the tensor with fixed bond dimension $D$ that best approximates the wavefunction made of updated tensors. Such an optimization is therefore non-local and requires the knowledge of the environment. In the simple update scheme, the environment is approximated by a set of diagonal weights $(\lambda^b_i)_{1 \leq i \leq D}$ on each bond $b$, which is unfortunately not well controlled. However, previous benchmarks with the much heavier full update in the case of the Shastry-Sutherland model showed a very small difference for a computational cost strongly reduced~\cite{wietek_thermodynamic_2019}. The singular value decomposition of the tensor network with weights allows for a truncation and provides a new set of weights $\lambda^b_i$ on the updated bond (see Fig.~\ref{fig:SU_1st}).

For the next-nearest neighbor terms, since there is no direct bond between the two updated sites, the entanglement is carried by their coupling to common nearest neighbors,  and one of the common nearest-neighbor sites has to be used as an intermediate tensor during the update, as shown in Fig.~\ref{fig:SU_2nd}. After the local tensor network contraction, two successive SVDs are used to renormalize the two bonds connecting the intermediate site to the next-nearest neighbor sites independently. All involved tensors are updated, including that of the nearest-neighbor site that has been used as an intermediate tensor. To avoid symmetry loss, a gate $\exp(-\tau h_2 /2)$ is applied twice successvely using the two common nearest neighbors as intermediate tensors.
Following ref.~\cite{corboz_simulation_2010}, we reduce the update complexity by first splitting all involved site tensors into a constant part and an effective part.

To improve the code stability, we introduced a relative cutoff around $10^{-10}$ in the singular values. This only matters at very high temperature, when the systems goes from $D=1$ to $D=D_\text{max}$ and the singular values are small. We used an imaginary time step around $\tau \sim 10^{-3}$ in most of our simulations, going down to $\tau\sim 10^{-4}$ for our largest values of $J_2$ in order to postpone the occurrence of a singularity (see last section of this Supplemental Material).
The exact values of $\tau$ and the cutoff need to be adjusted so that the same SU(2) representation appears on all bonds of the unit cell. These representations are set at very high temperature, where the cutoff matters, and are very stable through the time evolution. As long as this condition is fulfilled, we checked that varying these parameters makes a negligible difference for the observables.

To obtain information directly on the simple update process without needing to run the expensive CTMRG algorithm, we define a bond entanglement entropy for each bond $b$ of the unit cell as
\begin{equation}
S_\text{ent}(b) = -\sum_i \lambda^b_i \ln  \lambda^b_i
\end{equation}

%%%%%%%%%%%%%%%%%%%%%%%%%%%%%%%%%%%%%%%%%%%%%%%%%%%%%%
\subsection{Symmetry implementation}
Numerical precision plays a crucial role in imaginary time evolution since any error at high temperature propagates to low temperature. To improve the imaginary time evolution and prevent non-physical results, implementing symmetries is a compulsory step. 

We first implemented the abelian U(1) symmetry, as  discussed in ref~\cite{bauer_implementing_2011, singh_tensor_2011}. 
For any bond of the tensor network of dimension $\alpha$, we associate a list of $\alpha$ integer values corresponding to the quantum number $2 S^z$. In group theory terms, a bond is a representation of U(1) that we decompose into 1-dimensional irreducible representations (irreps) labeled by integers. A U(1) symmetric tensor is a tensor invariant under the action of U(1), or a U(1) singlet. In an appropriate basis, a U(1) symmetric tensor $T$ of rank $n$ takes a block diagonal form, meaning that a coefficient at position $(i_1\dots i_n)$ can be non-zero only if the sum of the associated quantum numbers at position $i_k$ is equal to 0.

To take full advantage of this block diagonal form, we implicitly reshape tensor $T$ as a matrix by merging the $n_r$ first legs into ``rows'' and the $n-n_r$ last ones into ``columns''. We combine the quantum numbers according to group fusion rules, summing them in the case of U(1). To abide by group theory rules, we have to use a covariant representation on the rows and a contravariant representation on the columns, which means we need to take the change the sign of the column quantum numbers. By swapping rows and columns to obtain contiguous symmetry sectors, we obtain a block diagonal matrix where each block is associated with a different U(1) quantum number. Our tensor acquired a more complex structure, with different blocks and an additional parameter $n_r$, changing $n_r$ being similar to transposing tensor legs. Fortunately, we only need to store the dense blocks instead of the full tensor, which in our simulation accounts for a reduction by a factor more than 5 in memory use.

To contract two tensors $A$ and $B$, we need to put the contracted legs as  $A$ ``columns'' and $B$ ``rows'', then tensor contraction becomes a blockwise matrix product. Linear algebra operation such as singular value decomposition (SVD) are also made independently in each symmetry sectors. Each block being much smaller than the initial matrices, we get a huge performance gain and can access larger bond dimension in the CTMRG. We also improved numerical precision by preventing any mixing into independent symmetry sectors that would arise in dense operations.

As we mentioned in the main article, we encountered  difficulties with the simple update when implementing only U(1), with an unphysical breaking of both SU(2) and $C_{4v}$ leading to wrong results. To overcome this problem, we improved numerical precision by implementing the full SU(2) symmetry at the level of the tensor~\cite{singh_tensor_2012, schmoll_programming_2020}. Since SU(2) is non-abelian, this is much more challenging than the U(1) case. This also impose constraints on the bond dimension $D$, which has to abide by irrep decomposition  (see Tab. \ref{tab:virt_irreps}).

\begin{figure}
	\centering
	\includegraphics[width=1.0\linewidth]{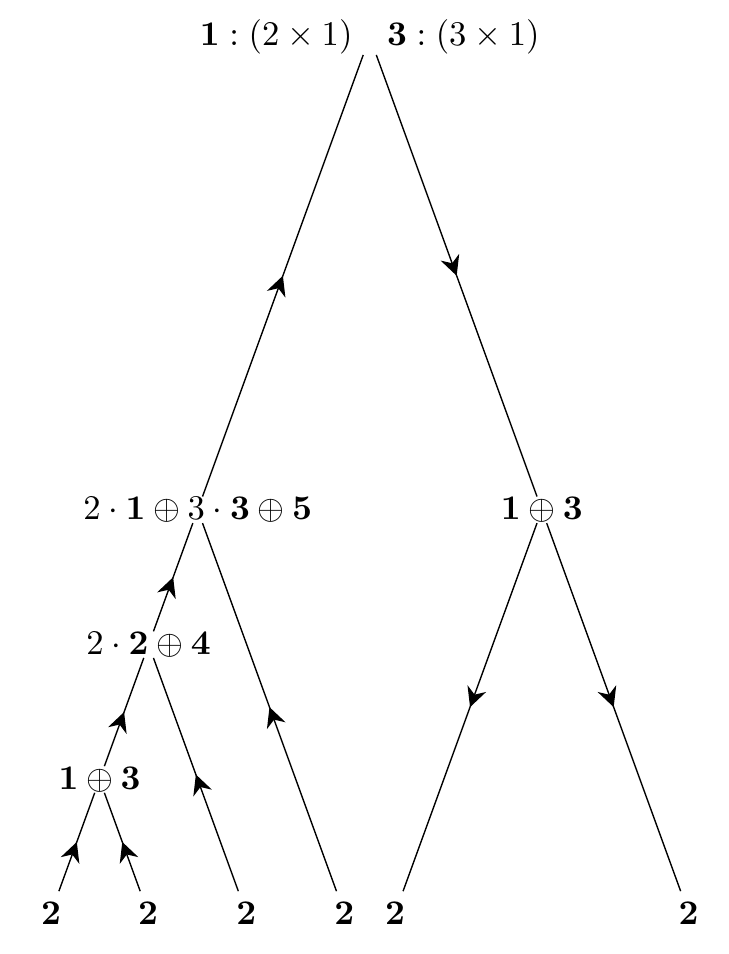}
	\caption{\footnotesize{Tree decomposition for a rank-6 SU(2) symmetric tensor, where each leg accounts for a spin 1/2 variable with dimension \textbf{2}.  Legs are split between $n_r = 4$ row legs (left, ascending arrows) and $6-n_r=2$ column legs (right, descending arrows) and fused into two separate trees, where only the first leg has depth. Each node corresponds to a Clebsch-Gordon tensor realizing the irrep decomposition of the two incoming legs. The output is a matrix with 2 blocks: one with shape $(2\times 1)$ associated with irrep \textbf{1} and one with shape $(3\times 1)$ associated with irrep \textbf{3}.}}
	\label{fig:su2_tree}
\end{figure}

The main ideas are similar: we reshape the tensor as a matrix, fuse ``row'' and ``column'' representations according to group theory rules, and decompose the matrix into independent blocks associated to one irrep. Then matrix products and linear algebra operations are all blockwise.
The main difference here is that this block diagonal matrix cannot be obtained by swapping coefficients and non-trivial projectors are required. We first need to specify a tree structure to fuse representation when merging legs: for simplicity, we impose a simple structure where only the first leg has depth (see Fig.~\ref{fig:su2_tree}). We then construct tree tensors of  Clebsch-Gordon coefficients realizing the irrep decomposition for each fusion.  At the top level, we contract the ``row'' tree with the ``column'' tree to obtain matrix blocks. This projector is then used to cast dense tensors into reduced, blockwise form as well as the reverse operation.

Transposing tensor legs (including changing $n_r$) is equivalent to a basis change of the total singlet space and requires unitary transformations. They are obtained by contracting together two tensors tree corresponding to the two different structures before and after transpose. While this operation is expensive, the resulting matrix can be efficiently stored as a sparse matrix and since it is a pure mathematical constant, it can be saved on the disk for latter use in independent simulations, regardless of any physical parameter.

\begin{table}
	\centering
	\begin{tabular}{CC}
		\hline
		\hline
		 D \quad & \quad V \\
		\hline
		1  \quad & \quad \textbf{1} \\
		4  \quad & \quad \textbf{1} \oplus \textbf{3} \\
		7  \quad & \quad \textbf{1} \oplus 2\cdot \textbf{3} \\
		8  \quad & \quad 2 \cdot \textbf{1} \oplus 2\cdot \textbf{3} \\
		11 \quad & \quad 2 \cdot \textbf{1} \oplus 3\cdot \textbf{3} \\
		16 \quad & \quad 2 \cdot \textbf{1} \oplus 3\cdot \textbf{3} \oplus \textbf{5}\\
		19 \quad & \quad 2 \cdot \textbf{1} \oplus 4\cdot \textbf{3} \oplus \textbf{5}\\
		22  \quad & \quad 2 \cdot \textbf{1} \oplus 5\cdot \textbf{3} \oplus \textbf{5}\\
		\hline
		\hline
	\end{tabular}
	\caption{Virtual space irrep decomposition for different values of $D$. When SU(2) symmetry is implemented, the virtual space on a bond is a representation of SU(2) and its dimension $D$ cannot be set arbitrary. In practice, the irrep decomposition is the same for all bonds and does not depend on $J_2$.}
	\label{tab:virt_irreps}
\end{table}

%%%%%%%%%%%%%%%%%%%%%%%%%%%%%%%%%%%%%%%%%%%%%%%%%%%%%%
\subsection{CTMRG and observables}
\begin{figure}
	\centering
	\includegraphics[width=1.0\linewidth]{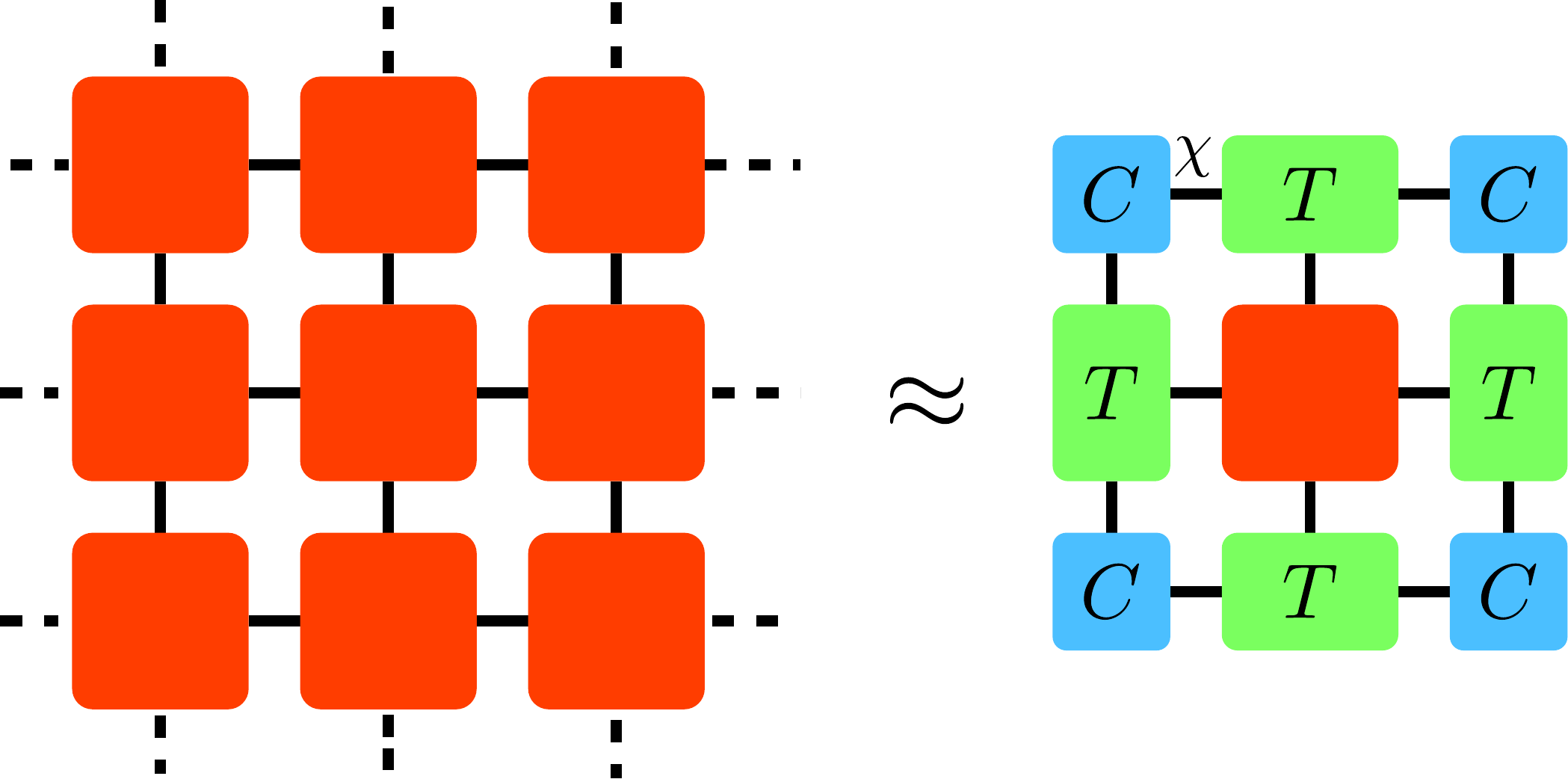}
	\caption{\footnotesize{CTMRG environment: an infinite environment is approximated by corners $C$ and edges $T$. The approximation is controlled by the dimension $\chi$ of the corner matrix.}}
	\label{fig:CTM}
\end{figure}

We define a bilayer tensor network by contracting a site tensor with its complex conjugate over their physical and ancilla leg. The remaining tensor only has virtual legs and can be used efficiently in contraction algorithms.
To contract the tensor network, we used the  Corner Transfer Matrix Renormalization Group (CTMRG) algorithm.  In the CTMRG, an infinite environment is approximated by a set of 4 corner tensors and 4 edge tensors (see Fig.~\ref{fig:CTM}). The accuracy of the approximation is controlled by the corner dimension $\chi$, taken as large as possible. In practice, we observe that the dependence of local observables (energy, Ising order parameter) on $\chi$ is small for the largest values of $\chi$ that are accessible. The main effect of having to use a finite $\chi$ is that, upon approaching the transition, the correlation length $\xi$ remains finite instead of diverging, an effect that has to be taken into account when locating the transition (see main text).

We implemented the asymmetric CTMRG algorithm proposed by Corboz, Rice and Troyer~\cite{corboz_competing_2014},  in which 4 inequivalent corners and 4 inequivalent edges are converged for all tensors in the unit cell. The exact setup is described in Ref.~\cite{corboz_competing_2014}. We just skipped the QR decomposition, which is not necessary to obtain valid projectors. We used a small cutoff of the order $10^{-10}$ in the projector singular values to improve numerical stability.

We only implemented the U(1) symmetry for the CTMRG, with an algorithm that dynamically selects the most relevant U(1) sectors in the singular value truncation. To avoid breaking SU(2) multiplets, we allowed for $\chi$ to fluctuate around a target value, always keeping full degenerate multiplets. We evaluated convergence with the Frobenius norm of two reduced density matrices in horizontal and vertical directions. This is much cheaper than evaluating the energy, which requires second neighbor density matrices.
To compute observables, we compute all inequivalent reduced density matrices in the unit cell for first and second neighbors from the CTMRG converged environment. We then take the average value inside the unit cell.

%%%%%%%%%%%%%%%%%%%%%%%%%%%%%%%%%%%%%%%%%%%%%%%%%%%%%%
\subsection{Transfer matrix and correlation length}
We approximate the horizontal and vertical transfer matrices using the CTMRG converged edge tensors. Due to the asymmetric CTMRG method, these matrices are not symmetric, however we checked that their spectra are always real up to numerical precision. We can define two different transfer matrices and two different spectra in both horizontal and vertical directions by shifting the unit cell from zero or one site. In practice the two sets of values are always very close along one given direction and we take the average.

The leading eigenvalue $\lambda_0$ is always a singlet. Then the other eigenvalues can be grouped by multiplicities, from which we infer their spin. For each of those degenerate values $\lambda_i$, we define a correlation length from the gap:
\begin{equation}
1/\xi_i = \log \abs{ \lambda_i / \lambda_0},
\end{equation}
Different eigenvalues are associated with different observables. We expect the leading triplet to be associated with magnetic order and the corresponding correlation length to diverge at zero temperature (long-range magnetic order), while close to the Ising phase transition the leading singlet is associated to the Ising order parameter and the correlation length associated to it diverges precisely at the transition.

%%%%%%%%%%%%%%%%%%%%%%%%%%%%%%%%%%%%%%%%%%%%%%%%%%%%%%
\section{Benchmark}
\begin{figure}
	\centering
	\includegraphics[width=1.0\linewidth]{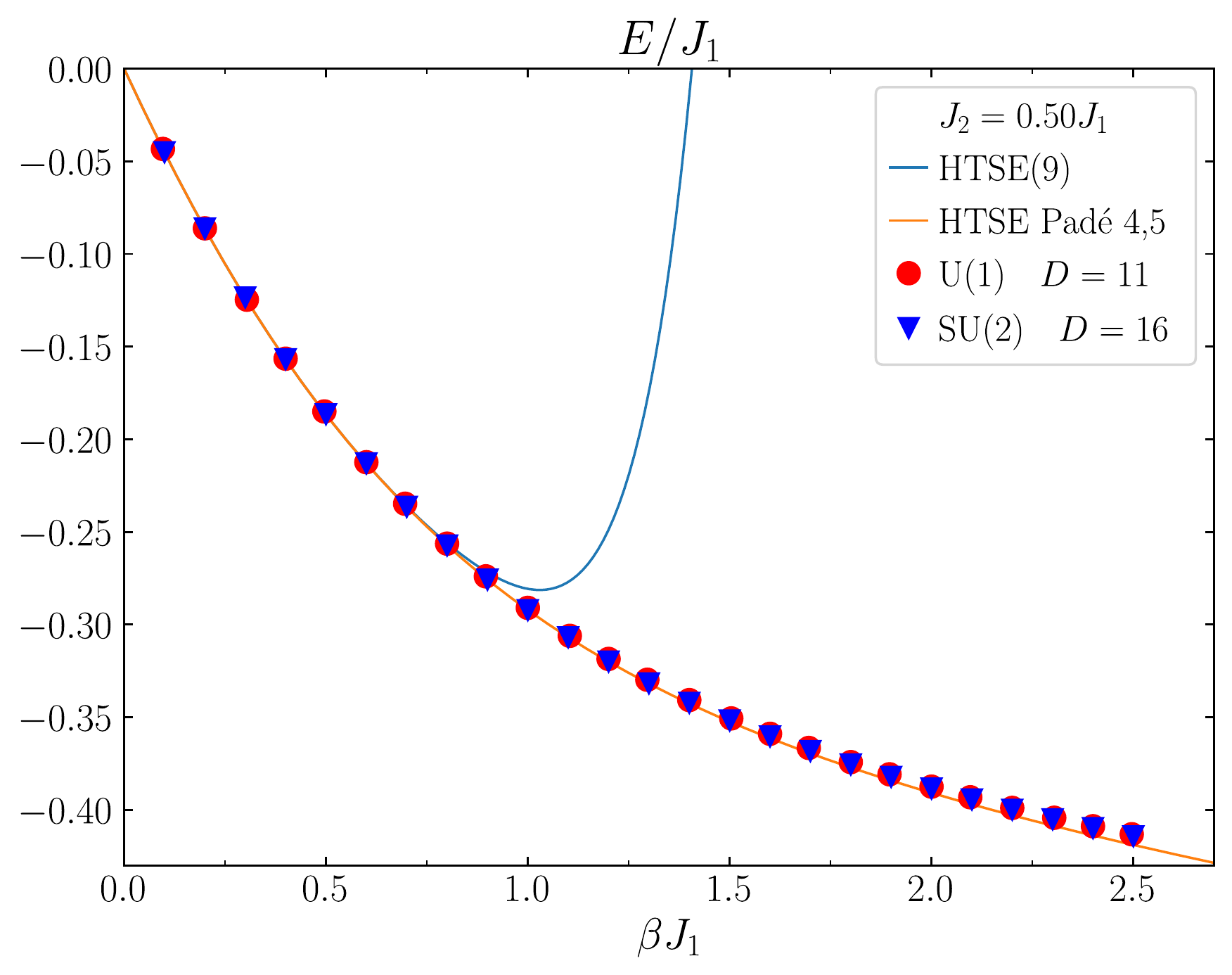}
	\caption{\footnotesize{Benchmark with high temperature series expansions and their $[4,5]$ Padé approximant for $J_2/J_1 = 0.50$. Results are converged in corner dimension $\chi$.}}
	\label{fig:benchmark_HTSE}
\end{figure}

We benchmarked our code with results obtained with high temperature series expansion~\cite{rosner_high-temperature_2003} at $J_2=0.5$. The results in Fig. \ref{fig:benchmark_HTSE} show perfect agreement with the Pad\'e approximant until high $\beta$ values. At such high temperatures, simulation parameters such as bond dimension $D$, imaginary time step $\tau$ or symmetry implementation have little impact on the energy value.

%%%%%%%%%%%%%%%%%%%%%%%%%%%%%%%%%%%%%%%%%%%%%%%%%%%%%%
\section{Simple update singularity}
By looking at the bond entanglement entropy, we obtain some information directly from the simple update. We can spot SU(2) symmetry breaking by checking multiplet degeneracies and explicit $C_{4v}$ symmetry breaking by comparing entropies on different bonds.

If we only implement U(1) symmetry, we observe that these entropies systematically undergo a dramatic change at some imaginary time value $\beta_\text{sing}$. The algorithm breaks SU(2) and, when $J_2 > J_{2c} \sim 0.6$, where the zero temperature phase has collinear order, this change also triggers a $C_{4v}$ symmetry breaking with a non-zero Ising order parameter.

While this singularity has some similarities with a phase transition, several clues show that this is a purely numerical artefact due to the lack of numerical precision: (i) SU(2) breaking is not small, while it is forbidden by Mermin-Wagner theorem; (ii) The exact value of $\beta_\text{sing}$  depends on $J_2$, but it also strongly depends on the value of the imaginary time step $\tau$, and more generally on purely computational simulation parameters such as float size; (iii) Instead of a narrow peak, the specific heat has a very broad maximum and does not go to zero at very low temperature; (iv) The correlation length does not diverge at $\beta = \beta_\text{sing}$.

We understand this artefact as a consequence of zero-temperature SU(2) symmetry breaking, which induces small deviations from perfect symmetry at finite temperature due to numerical error.

\begin{figure}
	\centering
	\includegraphics[width=1.0\linewidth]{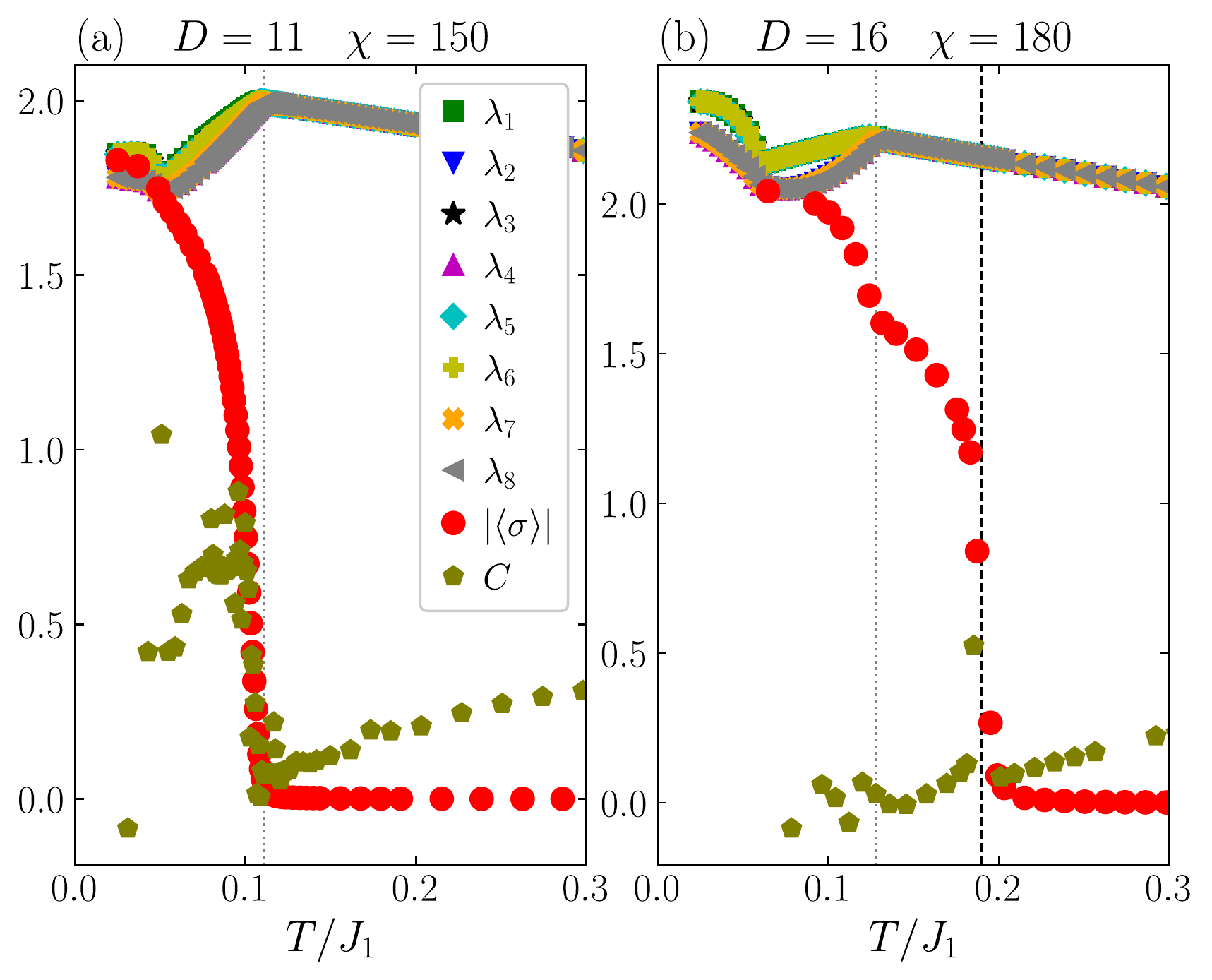}
	\caption{\footnotesize{Impact of simple update explicit symmetry breaking on observables at $J_2/J_1=1.20$. We plot the weight entropy for the 8 non-equivalent bonds of the unit cell $\lambda_i$ together with the absolute value of the Ising order parameter $\sigma$ and the specific heat $C$. (a) $D=11$: the simple update explicitly breaks $C_{4v}$, which can be seen in the bond entanglement entropy, where bonds have different entropies depending on their horizontal or vertical direction. While this is associated with a non-zero $\sigma$ and a surge in the specific heat, this is a pure artifact. (b) $D=16$: the phase transition arises around $T/J_1=0.19$, at a temperature where the simple update does not break explicitly $C_{4v}$ yet. It is associated with a very narrow specific heat peak. We then see the effect of the simple update artifact around $T/J_1 = 0.12$, which is similar to the $D=11$ case.}}
	\label{fig:obs_weights}
\end{figure}

By implementing SU(2), we completely removed the possibility of an SU(2) symmetry breaking and starkly improved numerical precision. However, for $J_2 > J_{2c}$, an explicit $C_{4v}$ still appears, with different bonds bearing different entanglement entropy. The characteristics of this singularity are very similar to the previous case, SU(2) breaking set aside. Fortunately, it only appears at much lower temperatures and we can spot the actual Ising phase transition, which has a very different signature and no explicit symmetry breaking on the simple-update bonds (see Fig.~\ref{fig:obs_weights}).

\bibliography{heisenberg_J1J2}